\documentclass[iop]{emulateapj}
\pdfoutput=1

\usepackage{epsf}
\usepackage{graphicx}
\usepackage{subfigure}
\usepackage{epstopdf}
\usepackage{apjfonts}
\usepackage{aas_macros}
\usepackage{amsmath}
\usepackage{amssymb}

\begin{document}
\title{MODEL-INDEPENDENT ESTIMATIONS FOR THE COSMIC CURVATURE FROM THE LATEST STRONG GRAVITATIONAL LENS SYSTEMS}

\author{HUAN ZHOU, ZHENGXIANG LI}

\affiliation{Department of Astronomy, Beijing Normal University, Beijing 100875, China: zxli918@bnu.edu.cn \\}

\begin{abstract}
Model-independent measurements for the cosmic spatial curvature, which is related to the nature of cosmic space-time geometry, plays an important role in cosmology. On the basis of the Distance Sum Rule in the Friedmann-Lema{\^i}tre-Robertson-Walker metric, (distance ratio) measurements of strong gravitational lensing (SGL) systems together with distances from type Ia supernovae observations have been proposed to directly estimate the spatial curvature without any assumptions for the theories of gravity and contents of the universe. However, previous studies indicated that a spatially closed universe was strongly preferred. In this paper, we re-estimate the cosmic curvature with the latest SGL data which includes 163 well-measured systems. In addition, possible factors, e.g. combination of SGL data from different surveys and stellar mass of the lens galaxy, which might affect estimations for the spatial curvature, are considered in our analysis. We find that, except the case where only SGL systems from the Sloan Lens ACS Survey are considered, a spatially flat universe is consistently favored at very high confidence level by the latest observations. It is suggested that the increasing number of well-measured strong lensing events might significantly reduce the bias of estimation for the cosmic curvature.

\end{abstract}


\section{Introduction}
If the universe satisfies the cosmological principle, i.e. the space of the universe is homogeneous and isotropic at large scales, we can use the Friedmann-Lema{\^i}tre-RobertsonWalker (FLRW) metric to describe the space-time geometry of the universe. In the the standard $\Lambda$CDM model, a spatially flat universe is favored at very high confidence level by several popular observations including the latest Planck-2015 results of cosmic microwave background (CMB) observations~\citep{Ade2016}. However, almost all of these constraints on the cosmic curvature were indirectly obtained by assuming some specific dark energy models. It should be noted that, on one hand, spatial curvature is related to the nature of space-time geometry. On the other hand, it also relates to the evolution of the universe and the various states of the universe in many cosmological models. For example, there is a strong degeneracy between the equation of state of dark energy and the spatial curvature in some standard dark energy scenarios~\citep{Clarkson2007, Gong2007}. Therefore, on the basis of the fundamental cosmological principle assumption, model-independent measurements for the spatial curvature have always been an important task in cosmology. That is, model-independent measurements for the spatial curvature is helpful for breaking this degeneracy and thus play an important role in exploring the nature of dark energy. In addition, small changes of the spatial curvature have a huge impact on the early inflation models~\citep{Eisenstein2005, Tegmark2006, Zhao2007, Wright2007}.

Recently, Clarkson et al. (2008) proposed to test the radial homogeneity of the universe in a model-independent way by directly measuring the spatial curvature at different redshifts with observations of the expansion rate and distance. Later, this test has been widely implemented with updated observations~\citep{Shafieloo2010, Mortsell2011, Sapone2014, Li2014, Cai2016}. Results consistently suggested that the cosmological principle is valid. Moreover, in this way, one can simultaneously achieve model-independent estimations for the spatial curvature. However, derivative of distance with respect to redshift $z$ usually leads to a large uncertainty. Therefore, dodging the derivative of distance with respect to redshift has been recently proposed to obtain constraints on the curvature with greater precision from observations of expansion rate and distance~\citep{Yu2016, Li2016, Wei2017}. Meanwhile, a similar test has also been presented to check the validity of the FLRW metric by using parallax distances and angular diameter distances~\citep{Rasanen2014}. Parallelly, the distance sum rule (DSR), which characterizes the relation of distances along null geodesics in the FLRW background, has been proposed to be a practical measurement of the curvature of the universe by studying the cross-correlation between foreground mass and gravitational shear of background galaxies~\citep{Bernstein2006}. More recently, DSR has been put forward as a consistency test~\citep{Rasanen2015}. That is, on one hand, the FLRW space-time will be ruled out if DSR is disfavored by observations; on the other hand, if observations are in well agreement with DSR, this test achieves a model-independent estimation for the spatial curvature of the universe. In R{\"a}s{\"a}nen et al. (2015), by using strong gravitational lensing (SGL) data selected from the Sloan Lens ACS Survey (SLACS) ~\citep{Bolton2008a} and the Union2.1 compilation of type Ia supernova (SNe Ia)~\citep{Suzuki2012}. They obtained that the spatial curvature parameter is constrained to be $\Omega_k=-0.55^{+1.18}_{-0.67}$  at $95\%$ confidence level, which slightly favors a spatially closed universe. So far, several following studies have been carried out with the latest SGL~\citep{Cao2015} and SNe Ia~\citep{Betoule2014} observations at that time~\citep{Xia2017, Li2018a}. In addition, Qi et al. (2018) have used radio quasars as distance indicators~\citep{Gurvtis1999, Cao2017a} together with SGL observations to extend the analysis to higher redshift. They obtained that, compared with constraints on the curvature obtained in~\citet{Rasanen2015}, a spatially flat universe is not preferred at higher confidence level. In this paper, following the DSR method, we re-estimate the spatial curvature with the latest 163 well-measured strong gravitational lensing systems~\citep{Cao2015, Shu2017, Chen2019}.

This paper is organized as follows. First, we will briefly introduce the distance sum rule and describe the observational data in Sec. \ref{sec2}. Next, in Sec. \ref{sec3}, we show corresponding constraint results. Finally, conclusions and discussions are presented in Sec. \ref{sec4}.

\section{Methodology}\label{sec2}
Following the assumption that cosmological principle is satisfied, the FLRW metric can be used to describe the space-time geometry of the universe (in units where c=1):
\begin{equation}\label{eq1}
ds^2=-dt^2+a^2(t)\bigg(\frac{dr^2}{1-kr^2}+r^2d\Omega^2\bigg),
\end{equation}
where $k$ is a constant related to the spatial curvature which takes the value -1, 0, or 1. Let $d_A(z_l, z_s)$ be the angular diameter distance of a source at redshift $z_s$ (emission time $t_s$) as seen at redshift $z_l$. Then we can define a dimensionless comoving angular diameter distance:
\begin{equation}\label{eq2}
d(z_l, z_s)\equiv(1+z_s)H_0d_A(z_l, z_s)=\frac{1}{\sqrt{\mid\Omega_k\mid}}{\Huge f}\bigg(\sqrt{\mid\Omega_k\mid}\int_{z_l}^{z_s}\frac{dz}{E(z)}\bigg),
\end{equation}
where
\begin{equation}\label{eq3}
f(x)=\left\{
   \begin{aligned}
   sin(x)\qquad\Omega_k<0, \\
   x\qquad\Omega_k=0, \\
   sinh(x)\qquad\Omega_k>0, \\
   \end{aligned}
   \right.
\end{equation}
$\Omega_k\equiv-k/(H_0^2a_0^2)$ ($a_0 = a(0)$ and $H_0$ are the present values of the scale factor and the Hubble parameter $H = \dot{a}/a$, respectively), and $E(z)\equiv H(z)/H_0$. In addition, we define $d_l$, $d_s$ and $d_{ls}$ as $d_l\equiv d(0, z_l)$, $d_s\equiv d(0, z_s)$, $d_{ls}\equiv d(z_l, z_s)$, respectively. Qualitatively, in the FLRW metric, $d_s =d_l+d_{ls}$  if the universe is spatially flat ($\Omega_k=0$). Meanwhile, $d_s >d_l+d_{ls}$ and $d_s <d_l+d_{ls}$ if the universe is spatially closed ($\Omega_k<0$) and open ($\Omega_k>0$), respectively. Quantitatively, these distances in the FLRW frame are related via
\begin{equation}\label{eq4}
\frac{d_{ls}}{d_s}=\sqrt{1+\Omega_kd_l^2}-\frac{d_l}{d_s}\sqrt{1+\Omega_kd_s^2}.
\end{equation}
This relation is a sum rule of distances in the FLRW universe and it was first proposed to obtain a model-independent estimate of the spatial curvature of in the case $|\Omega_k|\ll1$~\citep{Bernstein2006}. This sum rule is powerful since, in principle,
the validity of the FLRW metric can be tested by measuring three quantities ($d_l$, $d_s$ and $d_{ls}$)  for two different values ($z_l$ and $z_s$). Moreover, Equation 4 is also very general because it only assumes the geometrical optics and that ligh propagation is described with the FLRW metric. Different from the method between distance and expansion~\citep{Clarkson2007} , Equation 4 does not involve derivatives of the distance with respect to redshift. Unlike the method between angular diameter and parallax distances~\citep{Rasanen2014}, measurements of the distances on cosmological scales involved, $d_l$, $d_s$ and $d_{ls}$, are already currently available.

\subsection{Distances: $d_l$ and $d_s$}
In order to get model-independent estimates of the spatial curvature via Equation 4, we can, in principle, use different
kinds of distance indicators, such as standard candles, rulers, and sirens for providing distances $d_l$ and $d_s$ in the right-hand side of the Equation 4. Here, we use SNe Ia and intermediate-luminosity quasars (ILQSO) observations to obtain $d_l$ and $d_s$.

(i)SNe Ia data: In previous works~\citep{Xia2017, Li2018a}, the sample from a joint light-curve analysis (JLA) of SNe Ia observations obtained by the SDSS-II and SNLS collaborations~\citep{Betoule2014} is considered. The dataset includes several low redshift samples ($z < 0.1$), all three seasons from the SDSS-II ($0.05 < z < 0.4$), and three years from SNLS ($0.2 < z < 1$), and it totals 740 spectroscopically confirmed SNe Ia with high-quality light curves. For these well-measured events, the SALT2 model is used to reconstruct light-curve parameters ($x_1-$stretching of the light curve, $c-$the color of SNe Ia at maximum brightness, and $m^*_{\rm B}-$the observed peak magnitude in the rest-frame B band). In this case, the distance modulus $\mu=m^*_{\rm B} - M_{\rm B} + \eta*x_1 - \nu*c$  includes two nuisance parameters characterizing the stretch-luminosity and color-luminosity relationships. In other word, they represent the well-known broader-brighter and bluer-brighter relationships, respectively. The value of $M_{\rm B}$ is another nuisance parameter which denotes the absolute magnitude of a fiducial SNe. It was found that this the absolute magnitude of a fiducial SNe is dependent on the properties of host galaxies, e.g., the host stellar mass ($M_{\rm stellar}$). In the JLA SNe Ia, this dependence is approximately corrected with a simple step function when the mechanism has not been fully understood~\citep{Sullivan2011, Conley2011}.

Recently, \citet{Scolnic2018} released a new dataset called Pantheon SNe Ia, which consists of 1048 SNe Ia in the redshift range $0.01 < z < 2.3$. As the previously mentioned JLA SNe Ia, the nuisance parameters $\eta$ and $\nu$ are usually regarded as free parameters and are constrained together with cosmological parameters ~\citep{Li2011, Li2012}. However, this method might result in cosmological model dependence, therefore the distance calibrated in a specific cosmological model should not be directly used for other cosmological implications. To dodge this problem, \citet{Kessler2017} proposed a new method called BEAMS with Bias Corrections (BBC) to callibrated the SNe. This method is based on the approach proposed by \citet{Marriner2011} but includes extensive simulations for correcting the SALT2 light curve fitter. Moreover, the simulation also depends on an input cosmology, but the changes in the input cosmology within typical statistical uncertainties are in general negligible. With the BBC method, \citet{Scolnic2018} reported the corrected apparent magnitude $m^*_{B,corr} = m^*_B + \eta*x_1 - \nu*c +\Delta_B$ for all the SNe Ia. Therefore, we just need to subtract $M_B$ from $m^*_{B,corr}$ to calculate the distance moduli, relating to the luminosity distance $D_L$ via $\mu = 5log[\frac{D_L}{Mpc}]+25$ (please refer to \citep{Scolnic2018} for detailed information of the Pantheon SNe Ia).
\begin{equation}\label{eq5}
\mu^{\rm SN} = m^*_{\rm B,corr}-M_{\rm B}.
\end{equation}
Pantheon SNe Ia data has extended redshift to 2.3 and subtract some nuisance parameters, so we choose this as a distance indicator. With the distance-duality relation which holds in any space-time ~\citep{Etherington1933,Ellis2009}, the dimensionless comoving angular diameter distance $d=H_0D_L/(1+z)$ can be obtained by normalizing $H_0$. Here, the result from CMB measurements~\citep{Ade2016}, $H_0=67.74\pm0.46$, is used~\footnote{ We also checked the dependence of our analysis on $H_0$ by using the result from the latest local direct measurements \citep{Riess2019}, $H_0=74.03\pm1.42$, to normalize it, and found that this change leads to very tiny influence on estimations for cosmic curvature.}.

(ii)ILQSO data: Currently, the possibility of using compact radio sources to study cosmological parameters became very attractive~\citep{Zhu2002, Chen2003}. In order to extend our analysis to higher redshift, we use the sample of 120 ILQSO as distance indicators in the redshift range $0.46 < z < 2.76$~\citep{Cao2017a}. Our procedure follows the phenomenological model which quantifies the luminosity $L$ and redshift $z$ dependence of the linear sizes of quasars as~\citep{Gurvtis1994, Gurvtis1999, Cao2017a, Qi2018}
\begin{equation}\label{eq6}
l_m=lL^{\gamma}(1+z)^{n},
\end{equation}
where $l$ is the linear size scaling factor, $\gamma$ and $n$ quantify the dependence of the linear size on source luminosity and redshift, respectively. Following Cao et al. (2017a), for sample of 120 ILQSO, the linear size $l_m$ is independent of both redshift luminosity $L$ and redshift $z$ ($|\gamma|\simeq10^{-3}, |n|\simeq10^{-4}$), and the linear size $l_m$ was 11.03 pc at 2.29 GHz. We will use the above value of the linear size $l_m$ to calculate the angular diameter distances to ILQSO sample
\begin{equation}\label{eq7}
D_A(z)=\frac{l_m}{\theta(z)},
\end{equation}
where $\theta(z)$ is the angular size at redshift $z$ with error $\sigma_{\theta}$ ($\sigma_{\theta}^2=(\sigma_{ QSO}^{sta})^2+(\sigma_{QSO}^{sys})^2$, $\sigma_{QSO}^{sta}$ and $\sigma_{QSO}^{sys}$ are observational uncertainty and $10\%$ systematic uncertainty, respectively). Then one can use angular diameter distances to the quasars to obtain the dimensionless distances $d=H_0(1+z)D_A(z)$.

In our analysis, as proposed in R{\"a}s{\"a}nen et al. (2015), we model-independently determine the function of dimensionless angular diameter distance with respect to redshift (i.e., $d(z)$) by fitting a polynomial to SNe Ia data and ILQSO data. Here, we use a simple third-order polynomial function with initial conditions, $d(0)=0$ and $d'(0)=1$, to fit the cosmology-free distances of SNe Ia. This polynomial is expressed as~\footnote{ We also checked the dependence of our analysis on the form of $d(z)$ by using the logarithmic parametrization \citep{Risaliti2018} to fit the SNe Ia data and ILQSO data, and found that there is tiny difference between results when the polynomial expression and the logarithmic parametrization are considered separately.},
\begin{equation}\label{eq8}
d(z)=z+a_1z^2+a_2z^3,
\end{equation}
where $a_i$ are two free parameters which need to be constrained simultaneously with light-curve fitting parameters. It has been
suggested that, with current data, it does not make significant difference which function is used, as long as it is more flexible than a second order polynomial~\citep{Rasanen2015}. We show redshift distributions of both Pantheon SNe Ia samples, and ILQSO samples in Fig. \ref{Figure1}.

\subsection{The distance ratio: $d_{ls}/d_s$}
Since the discovery of the first gravitational lens Q0957+561~\citep{Walsh1979}, and with the increasing number of SGL systems detected, strong lensing has become an important astrophysical and cosmological probe~\citep{Futamase2001,Biesiada2006,Grillo2008}, e.g. tests of general relativity~\citep{Cao2017b}, tests of dark energy~\citep{Biesiada2010,Biesiada2011,Cao2012a,CaoZhu2014,Chen2015,Mario2019}, and constraints on cosmological models~\citep{CaoZhu2012, Cao2012b}. If general relativity holds on the scale of the lensing system, distance ratio $d_{ls}/d_s$ can be measured from observations for angular separation between lensed images of the same source and the structure of the lens. We will consider two types of lens models which have been extensively used in strong lensing studies.

(i)Singular isothermal ellipsoid (SIE) model: For SIE lens profile, we have
\begin{equation}\label{eq9}
\frac{d_{ls}}{d_s}=\frac{\theta_E}{4\pi\sigma^2f^2},
\end{equation}
where $\theta_E$ is the Einstein radius. In our analysis, we assign an error of $5\%$ on $\theta_{E}$ ~\citep{Cao2015}. $\sigma$ is the velocity dispersion of the lens, and $f$ is a phenomenological coefficient which characterizes uncertainty due to difference between the velocity dispersion of the observed stars and the underlying dark matter, and other systematic effects. In general, observations suggest the range $0.8 < f^2 < 1.2$ ~\citep{Ofek2003}. We take $f$ as a free parameter, which is assume as a flat prior, on the same weight as $\Omega_k$.~\citep{Li2018a}

(ii)Extended power-law (EPL) model: In addition to the SIE lens profile, we also consider a more complicate profile for the mass distribution of the lens, named as the Extended power-law (EPL) model. In this case, the distance ratio is written as ~\citep{Chen2019}
\begin{equation}\label{eq10}
\frac{d_{ls}}{d_s}=\frac{\theta_E}{4\pi\sigma^2f^2(\alpha,\beta,\delta)}\bigg(\frac{\theta}{\theta_E}\bigg)^{2-\alpha},
\end{equation}
where
\begin{equation}\label{eq11}
f(\alpha,\beta,\delta)=\sqrt{\frac{(\xi-2\beta)(3-\xi)\lambda(\alpha)\lambda(\delta)}{2\sqrt{\pi}(3-\delta)(\lambda(\xi)-\beta\lambda(\xi+2))}},
\end{equation}
with $\xi=\alpha+\delta-2$ and $\lambda(x)=\Gamma(\frac{x-1}{2})/\Gamma(\frac{x}{2})$. Moreover, we use two different kinds of velocity dispersion with different $\theta$. From the spectroscopic data, one can measure the velocity dispersion $\sigma_{ap}$ inside the circular aperture with the angular radius $\theta_{ap}$.  For the sake of comparison and in consideration of the effect of the aperture size on the measurements of velocity dispersions, all velocity dispersions $\sigma_{ap}$ measured within apertures of arbitrary sizes, are normalized to a typical physical aperture, $\sigma_{0}=\sigma_{ap}[\theta_{eff}/(2\theta_{ap})]^{\eta}$, with the radius $\theta_{eff}/2$, where $\theta_{eff}$ is the half-light radius of the lens galaxy. In this work, we adopt the best-fitting values of $\eta$ is $-0.066\pm0.035$ from ~\citet{Cappellari2006}. Then, the total uncertainty of $\sigma_{0}$ consists of the following three ingredients
\begin{equation}\label{eq12}
\big(\Delta\sigma_0^{{\rm tot}}\big)^2=\big(\Delta\sigma_0^{{\rm stat}}\big)^2+\big(\Delta\sigma_0^{{\rm AC}}\big)^2+\big(\Delta\sigma_0^{{\rm sys}}\big)^2,
\end{equation}
where $\Delta\sigma_0^{{\rm stat}}$ is the statistical error from the measurement error of $\sigma_{{\rm ap}}$. The error due to the aperture correction, $\Delta\sigma_0^{{\rm AC}}$ , is propagated from the uncertainty of $\eta$. In addition to the measurement errors, the uncertainty of $\sigma_{{\rm ap}}$ should include a systematic error of $3\%$, which mainly comes from the uncertainty in the projected gravitational mass $M_{grl}^{E}$~\citep{Jiang2007}. In addition, $\alpha$ and $\delta$ are free parameters which are from the total-mass density profile $\rho(r)\sim r^{-\alpha}$ and the luminosity density profile $\nu(r)\sim r^{-\delta}$, respectively. The parameter $\beta(r)=1-\sigma_{\theta}^2/\sigma_{r}^2$ denotes the anisotropy of the stellar velocity dispersion, and is also called as the stellar orbital anisotropy, where $\sigma_{\theta}$ and $\sigma_{r}$ are the tangential and radial velocity dispersions, respectively. In the literature, $\beta$ is usually assumed independent of $r$~\citep{Schwab2010, Cao2016, Xia2017, Qi2018, Chen2019}. Following these previous works, we also treat $\beta$ as a nuisance parameter and marginalize over it using a Gaussian prior with $\beta=0.18\pm0.13$, based on the well-studied sample of nearby elliptical galaxies.  In addition, we assume a flat prior for each remaining parameters of EPL model ($\alpha, \delta$).

The methodology described above is implemented to the sample of 163 galactic scale SGL systems from the Sloan Lens ACS Survey (SLACS)~\citep{Bolton2008a, Auger2009, Auger2010, Shu2015, Shu2017}, BOSS emission-line lens survey (BELLS)~\citep{Brownstein2012, Shu2016a, Shu2016b}, Lens Structure and Dynamics (LSD)~\citep{Koopmans2002, Koopmans2003, Treu2002, Treu2004} and Strong Lensing Legacy Survey (SL2S)~\citep{Ruff2011, Sonnenfeld2013a, Sonnenfeld2013b, Sonnenfeld2015} assembled by Chen et al. (2019). The SLACS is spectroscopic lens surveys in which candidates are selected from Sloan Digital Sky Survey (SDSS) data. Then candidates were followed up with Hubble Space Telescope (HST) photometry from Advanced Camera for Survey (ACS). The SDSS spectroscopy provides precise measurements, e.g. the stellar velocity dispersion and the redshifts of lens and source, while high-resolution HST imaging yields a detailed view of the lensed background source and the surface brightness profile of the lensing galaxy. BELLS is spectroscopic lens surveys in which candidates are selected from the Baryon Oscillation Spectroscopic Survey (BOSS) which has been initiated by upgrading SDSS-I optical spectrographs~\citep{Eisenstein2011}. Those candidates from SLACS and BELLS, which have multiple images or Einstein rings, have been classified as confirmed lenses. The LSD survey was a predecessor of SLACS by combining velocity dispersion data from ground-based (e.g. Keck) and photometric data from space-based (HST), which used different way compared with SLACS and BELLS to select lens systems. Therefor, in order to comply with SLACS and BELLS, there are just five reliable lenses from LSD~\citep{Cao2015, Chen2019}. Finally, the SL2S is a project dedicated to finding galaxy-scale lenses in the Canada-France-Hawaii-Telescope Legacy Survey. The targets are massive red galaxies which can be followed up with HST and spectroscopy.

\begin{figure}[htbp]
    \centering
     \includegraphics[width=0.49\textwidth, height=0.33\textwidth]{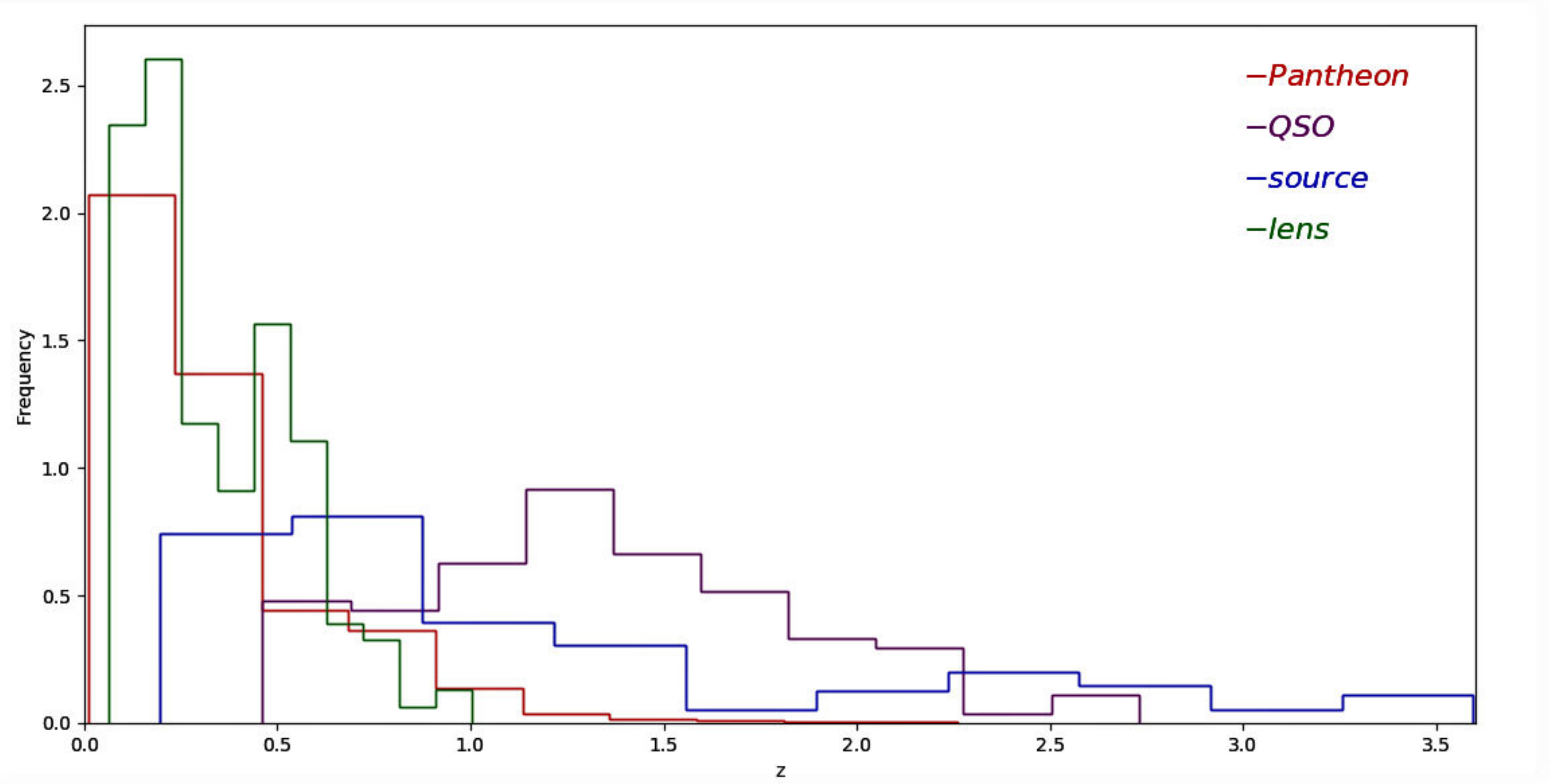}
	\caption{\label{Figure1}The redshift distributions for observational datasets including
sources and lenses from all 163 SGL samples, Pantheon SNe Ia, and ILQSO.}
\end{figure}

In Fig. \ref{Figure1}, we also show the redshift distribution of lens and source galaxies of SGL systems. A model-independent method of Gaussian processes~\citep{Seikel2012} was applied to reconstruct the dimensionless comoving distance from the ILQSO data straightforwardly, without any parametric assumption regarding cosmological model~\citep{Qi2018}. However, the fidelity of reconstructed function from most currently available observational data with the Gaussian processes should be further discussed and tested~\citep{Zhou2019}. Therefore, here, we combine Pantheon SNe Ia and ILQSO as distance indicators to extend redshift to 2.8. In this case, the maximum source redshift should be cut off at $z=2.8$. This full sample contains 152 SGL systems and is named as "Sample-152". Moreover, the stellar mass of lens galaxy also might result in possible bias in implications from SGL observations~\citep{Cao2012a, Cao2016, Xia2017, Li2018a}. Therefore, systems with typical velocity dispersion of the lens galaxy ranging from 200 to 300 $km\cdot s^{-1}$ were selected as a subsample for the sake of comparison. This subsample consists of 106 lensing systems and is name as "Sample-106". If we only use Pantheon SNe Ia as distance indicators, the maximum source redshift should be cut off at $z=2.3$. This subsample contains 137 SGL systems and is named as "Sample-137". To avoid possible systematic bias from combination of SGL data of different surveys, we estimate the cosmic curvature with the latest SGL data which only comes from the SLACS and includes 97 well-measured systems. This subsample is named as "Sample-97". Because the maximum source redshift of SLACS is 1.3, we just use Pantheon SNe Ia as distance indicators for "Sample-97".

\section{Results}\label{sec3}
As previously mentioned, the aim of this work is to re-estimate the spatial curvature with latest 163 strong gravitational lensing data. We infer the value of $\Omega_k$ via Eq. \ref{eq4} by confronting measurements of the latest distance indicators (Pantheon SNe Ia and ILQSO) with SGL observations. In our analysis, we perform a global fitting with the emcee ~\citep{Mackey2012}, using the Python module including Markov chain Monte Carlo. By marginalizing the light-curve fitting parameters ($M_{\rm B}$), and the polynomial coefficients ($a_1, a_2$), we obtain the marginalized distributions with $1\sigma$ and $2\sigma$ confidence levels contours for the curvature $\Omega_k$ and lens profile parameter(s) $f$ (or $\alpha, \beta$, and $\delta$).

\begin{figure}[htbp]
	\centering
	\includegraphics[scale=0.66]{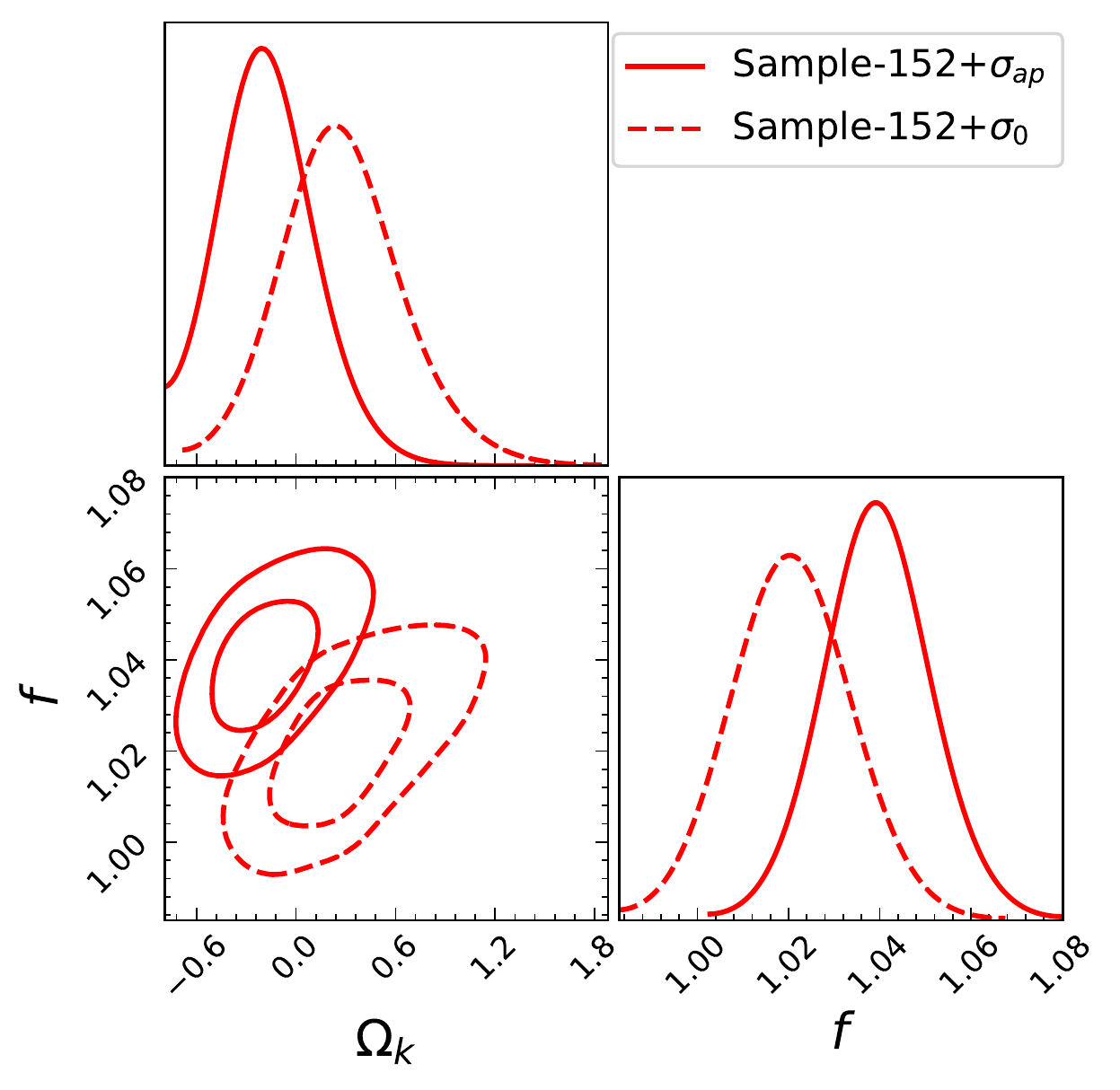}
	\caption{\label{Figure2}1D and 2D marginalized distributions with $1\sigma$ and $2\sigma$ confidence contours for the parameters $\Omega_k$ and $f$ constrained from the "Sample-152" by using the SIE model for lens mass profile. Solid line and dotted line are results when $\sigma_{ap}$ and $\sigma_0$ are used, respectively.}
\end{figure}

\begin{figure}[htbp]
	\centering
	\includegraphics[scale=0.66]{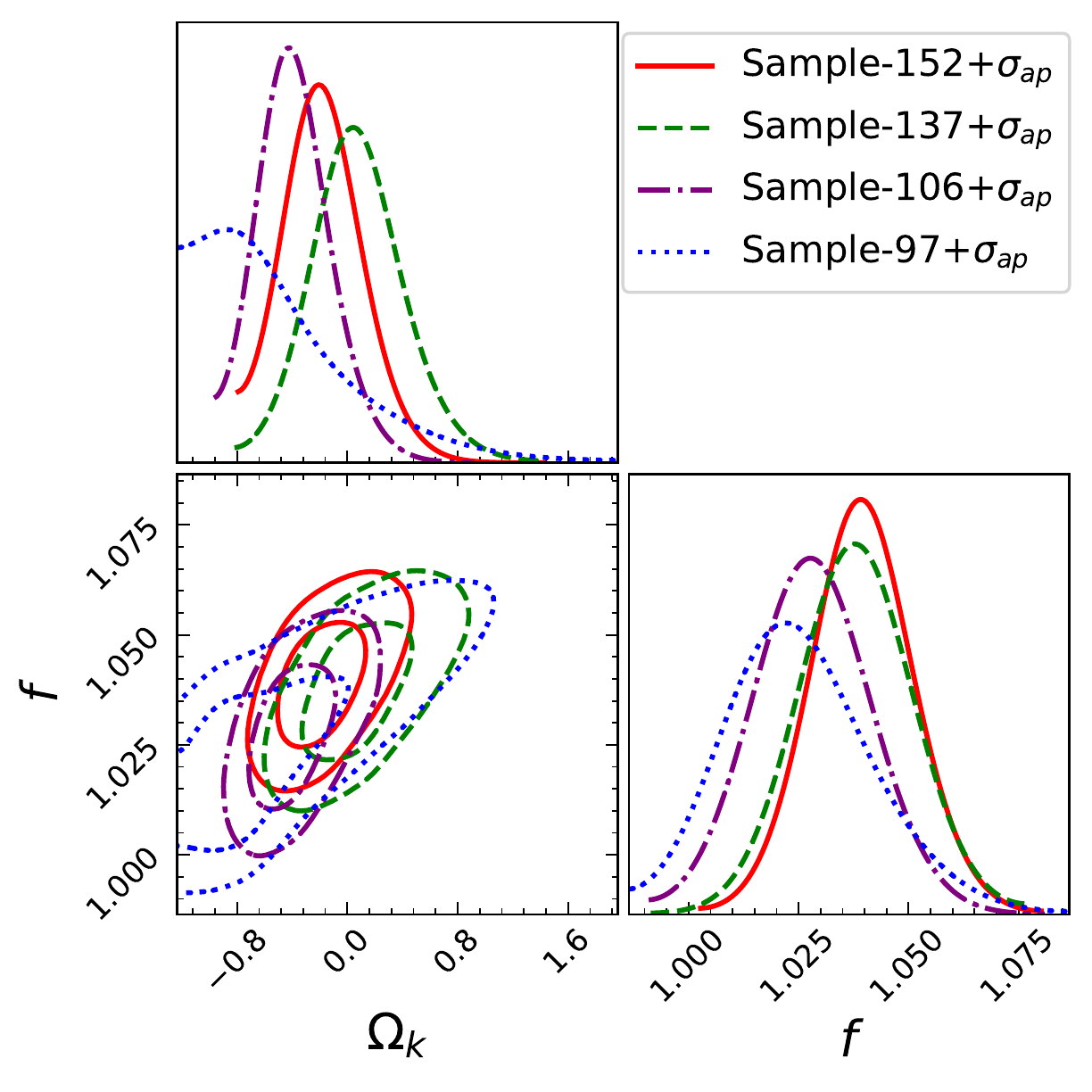}
	\caption{\label{Figure3}1D and 2D marginalized distributions with $1\sigma$ and $2\sigma$ confidence contours for the parameters $\Omega_k$ and $f$ that constrained from four sub-samples by using the SIE model for lens mass profile when $\sigma_{ap}$ is used.}
\end{figure}

\begin{figure}[htbp]
	\centering
	\includegraphics[scale=0.66]{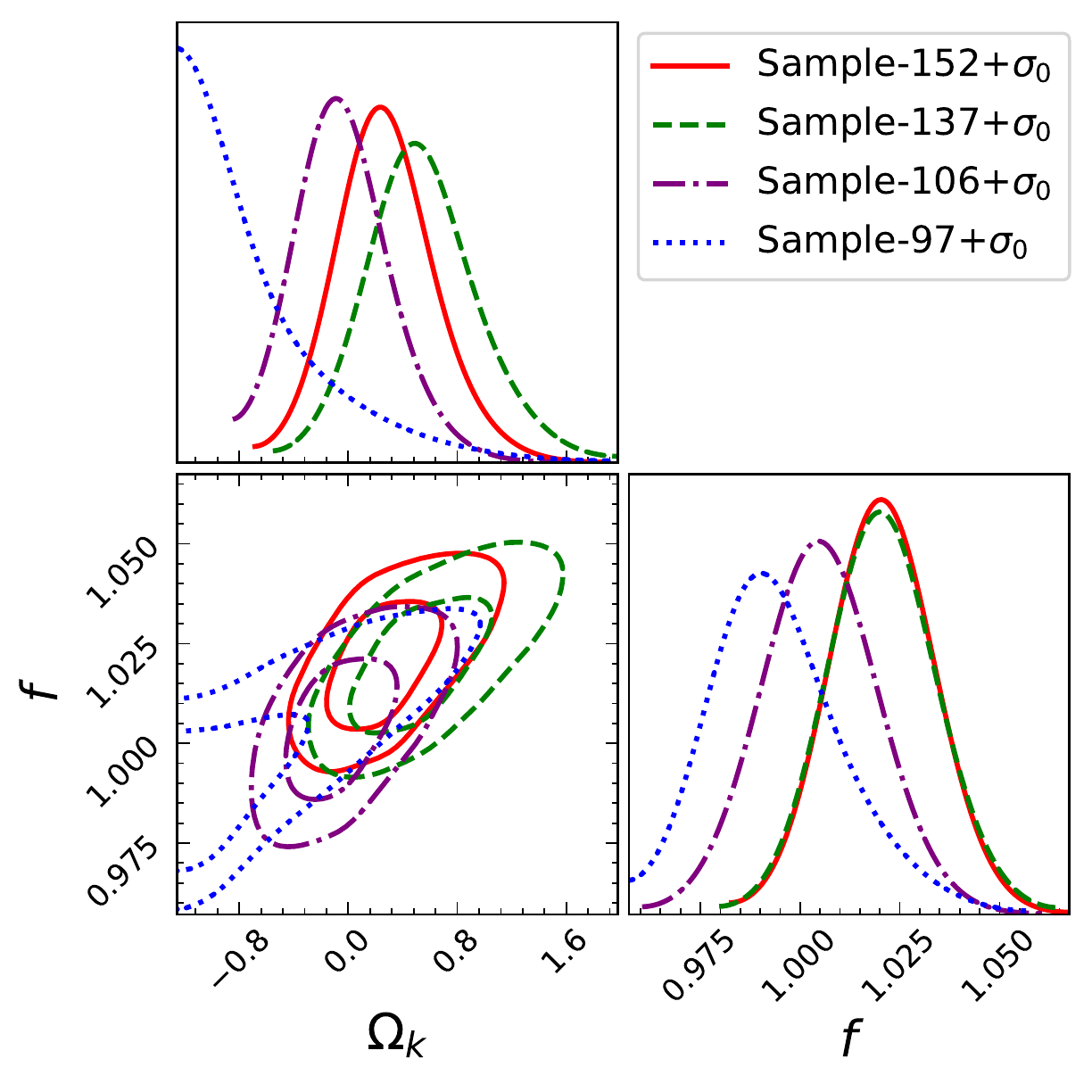}
	\caption{\label{Figure4}Same as in Figure 3, except using $\sigma_0$.}
\end{figure}

(i)SIE model: For the simple SIE model, graphical and numerical results from each SGL dataset are shown in Figs. (\ref{Figure2}-\ref{Figure4}) and Tab. \ref{table1}. For the full sample (Sample-152), $\Omega_k=-0.223^{+0.511}_{-0.305}$ and $\Omega_k=0.224^{+0.698}_{-0.428}$ are obtained at $95\%$ confidence level when $\sigma_{ap}$ and $\sigma_0$ are used, respectively. It is suggested that, compared to results in \citet{Rasanen2015}($\Omega_k=-0.55^{+1.18}_{-0.67}$ at $95\%$ confidence level) and \citet{Xia2017}($\Omega_k<0.6$ at $95\%$ confidence level), the precision of constraints on the $\Omega_k$ have been significantly improved due to the increase of the number of well-measured strong lensing systems. In addition, the latest distance indicators (Pantheon SNe Ia and ILQSO) extend the redshift of $d_l$ or $d_s$ data from 1.3 to 2.8, which allows us to add about 40 strong gravitational lens systems in this redshift range. It should be stressed that, in our analysis, we estimate all free parameters in a global fit without taking any priors for both $\Omega_k$ and $f$ into consideration. Moreover, we find that a spatially flat universe is consistently favored by these observations at high confidence levels. This is somewhat different from that obtained in \citet{Rasanen2015} where a trend of spatially closed universe has already been slightly indicated. Interestingly, these model-independent estimations are in good agreement with the conclusion of the latest CMB observations~\citep{Ade2016}. In addition, as shown in Fig. \ref{Figure2}, there is small discrepancy between the estimations of $f$ ($f=1.039^{+0.019}_{-0.017}$ and $f=1.020^{+0.020}_{-0.019}$ at $95\%$ confidence level) when $\sigma_{ap}$ and $\sigma_0$ are used separately. Results regarding $f$ suggest that the mass distribution profiles of lens galaxies are marginally consistent with the simplest singular isothermal sphere (SIS) model ($f =1$). These results are similar to what obtained in \citet{Xia2017} where the SIS profile is almost disfavored at more than $95\%$ confidence level when the prior for the curvature from the Planck 2015 CMB observations ($\Omega_k>-0.1$) is considered. On one hand, this discrepancy might imply that the SIS profile is too simple to characterize the mass distribution of lens galaxies. On the other hand, the deviations of constraints on $f$ from the SIS model also might indicate that lensing systems dominated by groups/clusters would affect fitting results \citep{Faure2011}. Last but not least, as shown in Figs. (\ref{Figure2}-\ref{Figure4}) and Tab. \ref{table1}, there are obvious differences in both cosmic spatial curvature and lens model parameters when $\sigma_{ap}$ and $\sigma_0$ are used separately.

\begin{figure}[htbp]
	\centering
	\includegraphics[scale=0.36]{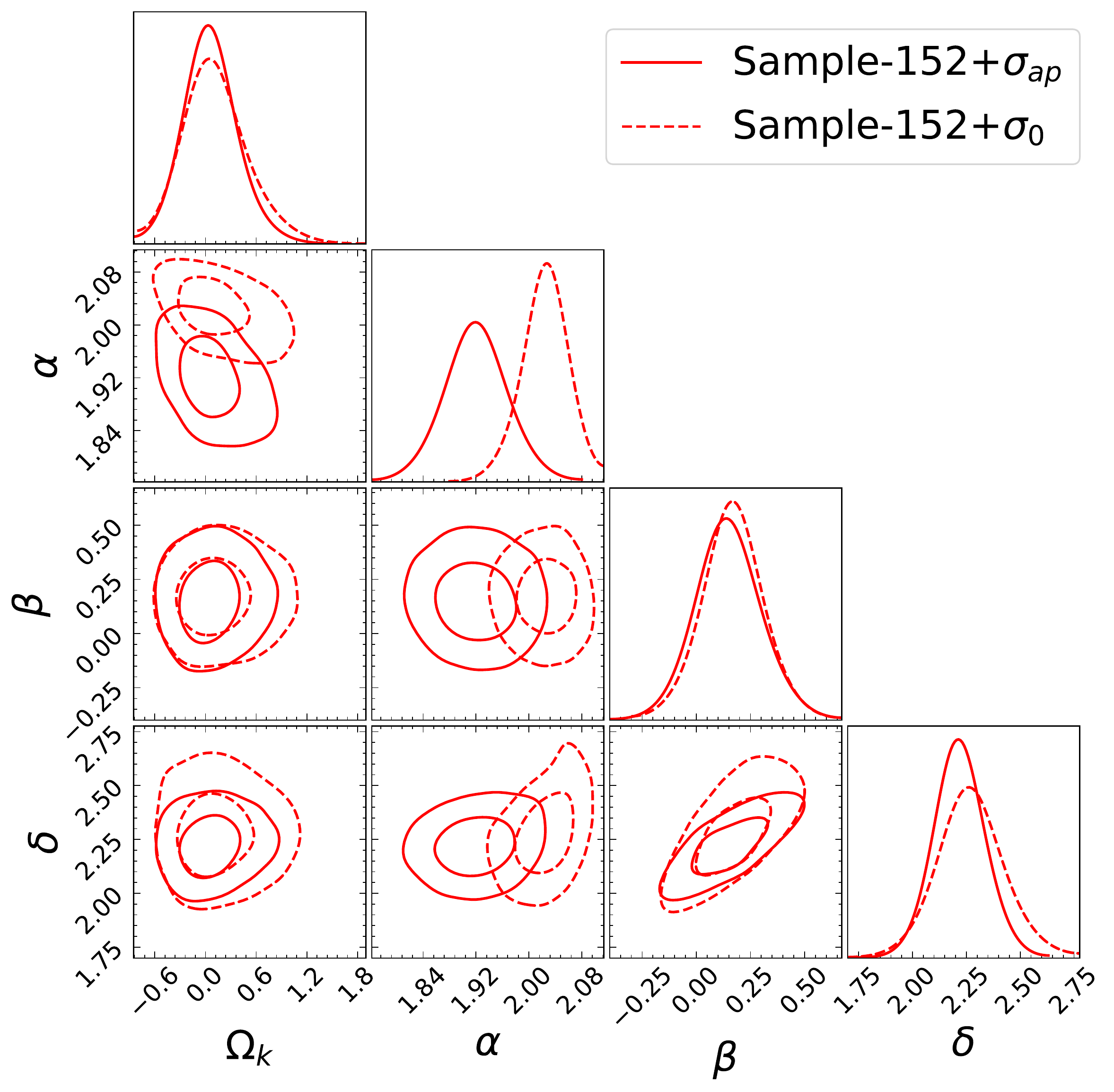}
	\caption{\label{Figure5}1D and 2D marginalized distributions with $1\sigma$ and $2\sigma$ confidence contours for the parameters $\Omega_k$, $\alpha$, $\beta$ and $\delta$ constrained from the "Sample-152" by using the EPL model for lens mass profile. Solid line and dotted line are results when $\sigma_{ap}$ and $\sigma_0$ are used, respectively.}
\end{figure}

\begin{figure}[htbp]
	\centering
	\includegraphics[scale=0.36]{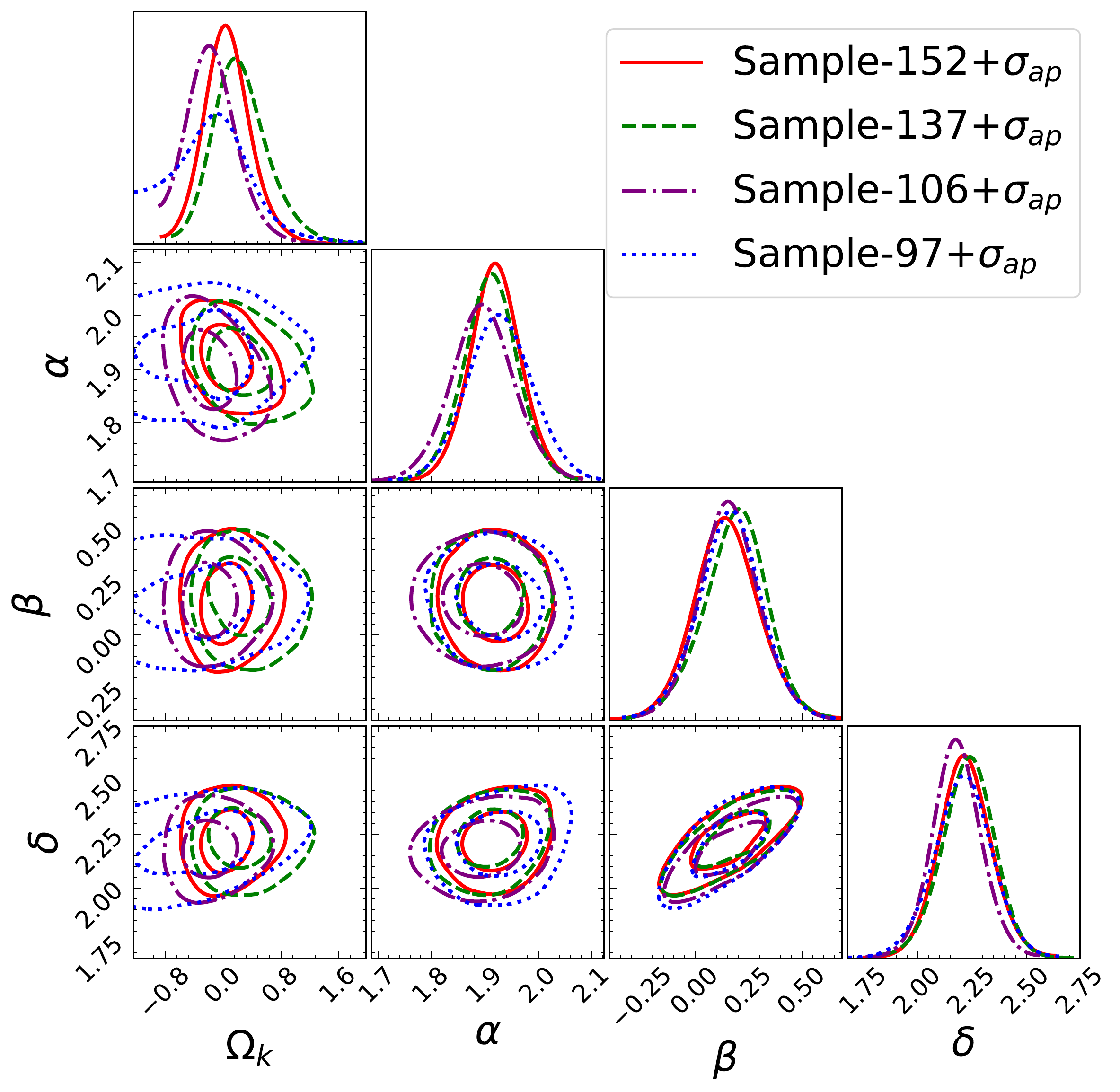}
	\caption{\label{Figure6}1D and 2D marginalized distributions with $1\sigma$ and $2\sigma$ confidence contours for the parameters $\Omega_k$, $\alpha$, $\beta$ and $\delta$ constrained from four sub-samples by using the EPL model for lens mass profile when $\sigma_{ap}$ are used.}
\end{figure}

\begin{figure}[htbp]
	\centering
	\includegraphics[scale=0.36]{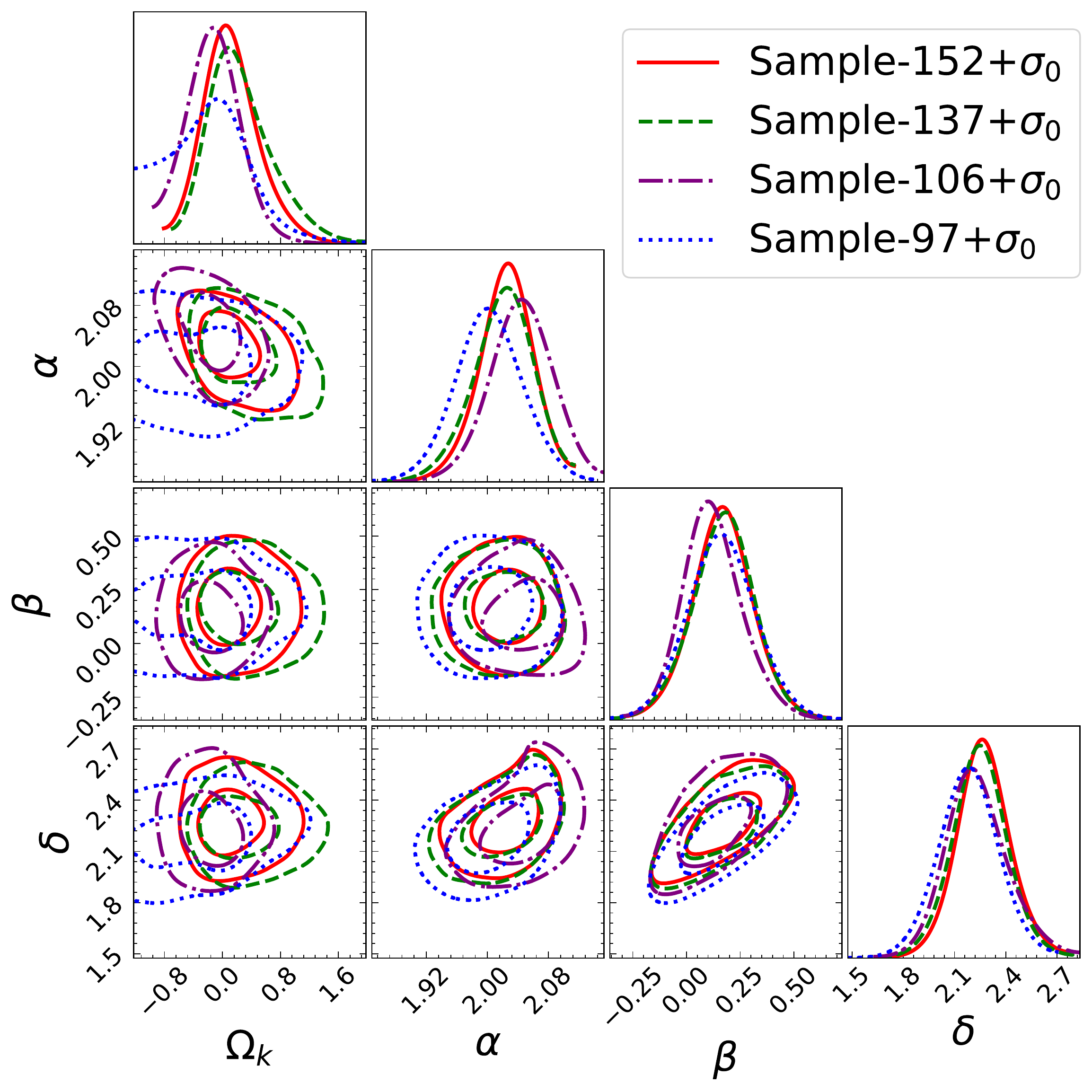}
	\caption{\label{Figure7}Same as in Figure 6, except using $\sigma_0$.}
\end{figure}

(ii)EPL model: For the EPL model, graphical and numerical results from each SGL dataset are shown in Figs. (\ref{Figure5}-\ref{Figure7}) and Tab. \ref{table2}. In the framework of the EPL model, comparing to results in SIE model, a spatially flat universe is more strongly supported by these available observations. In addition, as shown in Figs. (\ref{Figure5}) and Tab. \ref{table2}, the estimation of the spatial curvature is insensitive to different velocity dispersion, which means that the EPL model is more suitable to estimate the cosmic curvature. Previous observational constraints on the total-mass density profile was $\alpha\sim2$~\citep{Koopmans2009, Schwab2010, Sonnenfeld2013b, Oguri2014, Cao2016, Xia2017, Chen2019}. As shown in Figs. (\ref{Figure6}-\ref{Figure7}) and Tab. \ref{table2}, our constraints on it is marginally consistent with $\alpha=2$ at $95\%$ confidence level when $\sigma_{ap}$ is used. However, the constraints of the total-mass density profile are more consistent with $\alpha=2$, which means the velocity dispersion $\sigma_0$ is more suitable for the constraint of lens model parameters. In addition, the EPL model in which mass traces light ($\alpha=\delta=2$) is excluded at $>95\%$ confidence level, which might be helpful for understanding the difference in mass density distributions of dark matter and luminous baryons in early-type galaxies. As shown in Figs. (\ref{Figure5}-\ref{Figure7}), there are not only different degrees of degeneracy between parameters characterizing lens mass profile and $\Omega_k$, lens model parameters themselves also degenerate with each other. In this case, additional observational information, such as stellar velocity dispersion of the lens galaxy, can be possibly collected for providing complementary constraints on the slope of the total-mass density profile and thus are helpful for constraining $\alpha$, $\beta$, $\delta$, or even estimating $\Omega_k$. In addition, auxiliary data can be used to improve constraints on $\alpha$, $\beta$, and $\delta$ in the future~\citep{Cao2017b}. For example, $\alpha$ can be inferred for individual lenses from high resolution imaging of arcs~\citep{Suyu2007, Vegetti2010, Collett2014, Wong2015}, while constraints on $\beta$ and $\delta$ can be improved with integral field unit data~\citep{Barnabe2013}. Although the result of "Sample-97" can be compatible with a spatially flat universe at $95\%$ confidence level, it is obvious that the cosmic curvature of "Sample-97" is very different from other "Sample" in the SIE model. However, this differences are alleviated in the EPL model, which means that the EPL model is more reasonable to describe gravitational lens and estimate the cosmic curvature. On the whole, it is suggested that the EPL model with velocity dispersion $\sigma_0$ is more suitable for the constraints of parameters of both the lens and cosmic curvature.

It is suggested that constraints on cosmic curvature in the DSR method are not dependent on the different distance indicators
used and similar results are obtained. That is, model-independent estimations for curvature are mainly determined by SGL observations. Moreover, results derived from only the SLACS catalog, which is characterized by a selection function favoring moderately large-separation lenses \citep{Arneson2012}, suggest that this subsample might lead to biased estimation for the cosmic curvature. Therefore, more well-measured strong lensing events might be very useful for reducing the bias in estimation of curvature. In this sense, analyzing simulated SGL observations from Large Synoptic Survey Telescope (LSST) can give some predictions for constraints on cosmological parameters~\citep{Collett2015, Cao2018, Qi2018, Ma2019}. \citet{Qi2018} have found that, combining about 16000 strong lensing events combined with the distance information provided by 500 compact radio quasars, one can constrain the cosmic curvature with an accuracy of $\Omega_k\backsim10^{-3}$, which is comparable to the precision of Planck 2015 results. The upcoming LSST, which will monitor nearly half of the sky for 10 years by repeatedly scanning the field and is supposed to find $\thicksim10^{3}$ lensed quasars~\citep{Oguri2010}, will greatly improve this direct geometrical measurement for cosmic curvature.

\begin{table}
\small
\centering
\caption{Constraints on all parameters from the Pantheon  SNe Ia, ILQSO and SGL observations by
using the SIE model for lens mass profile.}\label{table1}
\scalebox{0.9}{
\begin{tabular}{ccccc}
\hline
parameters&Sample-152&Sample-137&Sample-106&Sample-97\\
\cline{1-5}
\multicolumn{5}{c}{$\sigma_{ap}$}\\
\cline{1-5}
$\Omega_k$&$-0.223_{-0.125}^{+0.231}$&$0.030_{-0.182}^{+0.287}$&$-0.428_{-0.141}^{+0.230}$&$-0.752_{-0.184}^{+0.701}$\\
$f$&$1.039_{-0.008}^{+0.009}$&$1.038_{-0.009}^{+0.010}$&$1.028_{-0.010}^{+0.010}$&$1.024_{-0.012}^{+0.015}$\\
$M_B$&$-19.383_{-0.006}^{+0.006}$&$-19.429_{-0.007}^{+0.007}$&$-19.383_{-0.006}^{+0.006}$&$-19.426_{-0.007}^{+0.007}$\\
$a_1$&$-0.296_{-0.008}^{+0.007}$&$-0.215_{-0.014}^{+0.013}$&$-0.296_{-0.008}^{+0.008}$&$-0.226_{-0.013}^{+0.013}$\\
$a_2$&$0.021_{-0.004}^{+0.004}$&$-0.007_{-0.011}^{+0.012}$&$0.021_{-0.005}^{+0.005}$&$0.004_{-0.011}^{+0.011}$\\
\cline{1-5}
\multicolumn{5}{c}{$\sigma_{0}$}\\
\cline{1-5}
$\Omega_k$&$0.224_{-0.184}^{+0.302}$&$0.483_{-0.239}^{+0.385}$&$-0.105_{-0.169}^{+0.311}$&$-0.969_{-0.204}^{+0.734}$\\
$f$&$1.020_{-0.010}^{+0.010}$&$1.020_{-0.010}^{+0.011}$&$1.005_{-0.010}^{+0.011}$&$0.992_{-0.010}^{+0.016}$\\
$M_B$&$-19.383_{-0.006}^{+0.006}$&$-19.429_{-0.007}^{+0.007}$&$-19.383_{-0.006}^{+0.006}$&$-19.426_{-0.007}^{+0.008}$\\
$a_1$&$-0.295_{-0.008}^{+0.008}$&$-0.216_{-0.014}^{+0.013}$&$-0.295_{-0.008}^{+0.008}$&$-0.226_{-0.014}^{+0.013}$\\
$a_2$&$0.020_{-0.005}^{+0.005}$&$-0.006_{-0.011}^{+0.011}$&$0.020_{-0.005}^{+0.005}$&$0.004_{-0.011}^{+0.011}$\\
\hline
\end{tabular}}
\end{table}

\begin{table}
\small
\centering
\caption{Constraints on all parameters from the Pantheon SNe Ia, ILQSO and SGL
observations by using the EPL model for lens mass profile.}\label{table2}
\scalebox{0.9}{
\begin{tabular}{ccccc}
\hline
parameters&Sample-152&Sample-137&Sample-106&Sample-97\\
\cline{1-5}
\multicolumn{5}{c}{$\sigma_{ap}$}\\
\cline{1-5}
$\Omega_k$&$0.024_{-0.150}^{+0.234}$&$0.138_{-0.076}^{+0.396}$&$-0.194_{-0.163}^{+0.230}$&$-0.053_{-0.691}^{+0.184}$\\
$\alpha$&$1.919_{-0.033}^{+0.036}$&$1.912_{-0.038}^{+0.033}$&$1.894_{-0.039}^{+0.045}$&$1.931_{-0.043}^{+0.048}$\\
$\delta$&$2.207_{-0.055}^{+0.092}$&$2.245_{-0.093}^{+0.047}$&$2.167_{-0.052}^{+0.090}$&$2.218_{-0.099}^{+0.068}$\\
$M_B$&$-19.382_{-0.006}^{+0.006}$&$-19.430_{-0.009}^{+0.008}$&$-19.383_{-0.006}^{+0.006}$&$-19.426_{-0.007}^{+0.007}$\\
$a_1$&$-0.296_{-0.008}^{+0.007}$&$-0.213_{-0.014}^{+0.012}$&$-0.295_{-0.008}^{+0.008}$&$-0.225_{-0.014}^{+0.013}$\\
$a_2$&$0.021_{-0.004}^{+0.005}$&$-0.010_{-0.008}^{+0.013}$&$0.020_{-0.005}^{+0.005}$&$0.004_{-0.012}^{+0.012}$\\
\cline{1-5}
\multicolumn{5}{c}{$\sigma_{0}$}\\
\cline{1-5}
$\Omega_k$&$0.012_{-0.093}^{+0.381}$&$0.100_{-0.114}^{+0.538}$&$-0.078_{-0.320}^{+0.118}$&$-0.048_{-0.772}^{+0.154}$\\
$\alpha$&$2.027_{-0.027}^{+0.024}$&$2.026_{-0.037}^{+0.023}$&$2.043_{-0.026}^{+0.034}$&$2.000_{-0.028}^{+0.036}$\\
$\delta$&$2.258_{-0.084}^{+0.122}$&$2.253_{-0.107}^{+0.088}$&$2.187_{-0.055}^{+0.185}$&$2.182_{-0.115}^{+0.118}$\\
$M_B$&$-19.383_{-0.006}^{+0.005}$&$-19.430_{-0.007}^{+0.007}$&$-19.383_{-0.006}^{+0.006}$&$-19.425_{-0.008}^{+0.008}$\\
$a_1$&$-0.295_{-0.008}^{+0.008}$&$-0.213_{-0.015}^{+0.011}$&$-0.295_{-0.008}^{+0.008}$&$-0.226_{-0.013}^{+0.013}$\\
$a_2$&$0.020_{-0.005}^{+0.005}$&$-0.009_{-0.010}^{+0.012}$&$0.021_{-0.005}^{+0.005}$&$0.004_{-0.011}^{+0.011}$\\
\hline
\end{tabular}}
\end{table}

\section{Conclusion}\label{sec4}
On the basis of the cosmological principle, we can use the FLRW metric to describe the space-time geometry of the universe. In this paper, by using the Distance Sum Rule in the FLRW metric, we obtain direct geometrical estimations for the spatial curvature of the universe. These estimations are independent of the energy-momentum contents of the universe and the validity of the Einstein equation on cosmological scales. We re-estimate the cosmic curvature by combining the latest SGL data which includes 163 well-measured systems with the latest distance indicators (Pantheon SNe Ia and ILQSO).  Along with the spatial curvature $\Omega_k$, parameters characterizing the mass profile of lens galaxies, and polynomial coefficients are simultaneously constrained in a global fitting. Graphic results on concerned parameters ($\Omega_k$, and $f$ or $\alpha, \beta, \delta$) are shown in Figs. (\ref{Figure2}-\ref{Figure7}) and estimations for all parameters are summarized in Tabs. (\ref{table1}-\ref{table2}). In summary, compared to the results obtained in \citet{Rasanen2015,Li2018a} where a spatially closed universe was preferred, the constraints on the spatial curvature from the latest SGL data in our analysis strongly favor a spatially flat universe. It is suggested that  more well-measured strong lensing systems together with good distance indicators might reduce the bias in these model-independent estimations. However, it should be pointed out that there still is a large gap between the precision of our model-independent estimations and that obtained from the Planck-2015 CMB observations in the standard $\Lambda$CDM model. Therefore, a large number of SGL systems from the program in the near future, e.g. the Euclid satellite and the LSST, are expected to obtain more precise constraints on the spatial curvature of universe. In addition, other more promising lensing systems have been proposed, for example, time delay measurements of strongly lensed transients (such as gravitational waves and fast radio bursts) as a precision probe to constrain some fundamental cosmological parameters including the spatial curvature~\citep{Fan2017, Liao2017a, Li2018b, Li2019}. These upcoming improvements on the precision of model-independent estimation of cosmic curvature will be of great significance for breaking the degeneracy between the curvature and dark energy, and thus will be very helpful for studying the nature of dark energy or even understanding the physical mechanism of cosmic acceleration.

\section{Acknowledgments}
We would like to thank Shuo Cao and Yun Chen for helpful discussions. We are also very grateful to the referee for his/her valuable comments which have allowed us to significantly improve our manuscript. This work was supported by the National Natural Science Foundation of China under Grants Nos. 11505008 and the Interdiscipline Research Funds of Beijing Normal University.


\begin{thebibliography}{0}%
\makeatletter
\providecommand \@ifxundefined [1]{%
 \@ifx{#1\undefined}
}%
\providecommand \@ifnum [1]{%
 \ifnum #1\expandafter \@firstoftwo
 \else \expandafter \@secondoftwo
 \fi
}%
\providecommand \@ifx [1]{%
 \ifx #1\expandafter \@firstoftwo
 \else \expandafter \@secondoftwo
 \fi
}%
\providecommand \natexlab [1]{#1}%
\providecommand \enquote  [1]{``#1''}%
\providecommand \bibnamefont  [1]{#1}%
\providecommand \bibfnamefont [1]{#1}%
\providecommand \citenamefont [1]{#1}%
\providecommand \href@noop [0]{\@secondoftwo}%
\providecommand \href [0]{\begingroup \@sanitize@url \@href}%
\providecommand \@href[1]{\@@startlink{#1}\@@href}%
\providecommand \@@href[1]{\endgroup#1\@@endlink}%
\providecommand \@sanitize@url [0]{\catcode `\\12\catcode `\$12\catcode
  `\&12\catcode `\#12\catcode `\^12\catcode `\_12\catcode `\%12\relax}%
\providecommand \@@startlink[1]{}%
\providecommand \@@endlink[0]{}%
\providecommand \url  [0]{\begingroup\@sanitize@url \@url }%
\providecommand \@url [1]{\endgroup\@href {#1}{\urlprefix }}%
\providecommand \urlprefix  [0]{URL }%
\providecommand \Eprint [0]{\href }%
\providecommand \doibase [0]{http://dx.doi.org/}%
\providecommand \selectlanguage [0]{\@gobble}%
\providecommand \bibinfo  [0]{\@secondoftwo}%
\providecommand \bibfield  [0]{\@secondoftwo}%
\providecommand \translation [1]{[#1]}%
\providecommand \BibitemOpen [0]{}%
\providecommand \bibitemStop [0]{}%
\providecommand \bibitemNoStop [0]{.\EOS\space}%
\providecommand \EOS [0]{\spacefactor3000\relax}%
\providecommand \BibitemShut  [1]{\csname bibitem#1\endcsname}%
\let\auto@bib@innerbib\@empty
\end{thebibliography}%


\begin{thebibliography}{}

\bibitem[Planck Collaboration et al. (2016)]{Ade2016}Ade, P. A. R., Aghanim, N., Arnaud, M., et al. (Planck Collaboration) 2016, A$\&$A, 594, A13

\bibitem[Arneson et al. (2012)]{Arneson2012}Arneson, R. A., Brownstein, J. R., \& Bolton, A. S. 2012, ApJ, 753, 4

\bibitem[Auger et al. (2009)]{Auger2009}Auger, M. W., Treu, T., Bolton, A. S., Gavazzi, R., et al. 2009, ApJ. 705, 1099

\bibitem[Auger et al. (2010)]{Auger2010}Auger, M. W., Treu, T., Bolton, A. S., et al. 2010, ApJ. 724, 511

\bibitem[Barnab\`e et al. (2013)]{Barnabe2013}Barnab{\`e}, M., et al. 2013, MNRAS, 436, 253

\bibitem[Bernstein (2006)]{Bernstein2006}Bernstein, G. 2006, ApJ, 637, 598

\bibitem[Betoule et al. (2014)]{Betoule2014}Betoule, M., Kessler, R., Guy, J., et al. 2014, A$\&$A, 568, A22

\bibitem[Biesiada (2006)]{Biesiada2006}Biesiada, M. 2006, PRD, 69, 101305

\bibitem[Biesiada et al. (2010)]{Biesiada2010}Biesiada, M., Pi{\'o}rkowska A., $\&$ Malec B. 2010, MNRAS, 406, 1055

\bibitem[Biesiada et al. (2011)]{Biesiada2011}Biesiada, M., Malec B., $\&$ Pi{\'o}rkowska A. 2011, RAA, 11, 641

\bibitem[Bolton et al. (2006)]{Bolton2006}Bolton, A. S., Burles, S., Koopmans, L. V. E., et al. 2006, ApJ, 638, 703

\bibitem[Bolton et al. (2008a)]{Bolton2008a}Bolton, A. S., Burles, S., Koopmans, L. V. E., et al. 2008, ApJ, 682, 964

\bibitem[Bolton et al. (2008b)]{Bolton2008b}Bolton, A. S., Burles, S., Koopmans, L. V. E., et al. 2008, ApJ, 684, 248

\bibitem[Brownstein et al. (2012)]{Brownstein2012}Brownstein, J. R., Bolton, A. S., Schlegel, D. J., et al. 2012, ApJ, 744, 41

\bibitem[Cai et al. (2016)]{Cai2016}Cai, R.-G., Guo, Z.-K., $\&$ Yang, T. 2016, PRD, 93, 043517

\bibitem[Cao \& Zhu (2012)]{CaoZhu2012}Cao, S., $\&$ Zhu, Z.-H. 2012, A$\&$A, 538, A43

\bibitem[Cao et al. (2012a)]{Cao2012a}Cao, S., Covone, G., $\&$  Zhu, Z.-H. 2012, ApJ, 755, 31

\bibitem[Cao et al. (2012b)]{Cao2012b}Cao, S., et al. 2012, JCAP, 03, 016

\bibitem[Cao \& Zhu. (2014)]{CaoZhu2014}Cao, S., $\&$ Zhu Z.-H. 2014, PRD, 90, 083006

\bibitem[Cao et al. (2015)]{Cao2015}Cao, S., Biesiada, M., Yao, M., Gavazzi, R., $\&$ Zhu, Z.-H. 2015, ApJ, 805, 185

\bibitem[Cao et al. (2016)]{Cao2016}Cao, S., Biesiada, M., Yao, M., $\&$ Zhu, Z.-H. 2016, MNRAS, 461, 2192

\bibitem[Cao et al. (2017a)]{Cao2017a}Cao, S., et al. 2017, A$\&$A, 606, A15

\bibitem[Cao et al. (2017b)]{Cao2017b}Cao, S., et al. 2017, ApJ, 835, 92

\bibitem[Cao et al. (2018)]{Cao2018}Cao, S., et al. 2018, ApJ, 867, 1

\bibitem[Cappellari et al. (2006)]{Cappellari2006}Cappellari, M., et al. 2006, MNRAS, 366, 1126

\bibitem[Chen \& Ratra (2003)]{Chen2003}Chen, G., $\&$ Ratra, B. 2003, ApJ, 582, 586

\bibitem[Chen et al. (2015)]{Chen2015}Chen, Y., Geng, C.-Q., Cao, S. et al. 2015, JCAP, 02, 010

\bibitem[Chen et al. (2019)]{Chen2019}Chen, Y., Li, R., $\&$ Shu, Y. 2019, MNRAS, 488, 2977

\bibitem[Clarkson, Cortes, \& Bassett (2007)]{Clarkson2007}Clarkson, C. Cortes, M., $\&$ Bassett, B.A. 2007, JCAP, 08, 011

\bibitem[Clarkson, Bassett, \& Lu (2008)]{Clarkson2008}Clarkson, C., Bassett, B. A., $\&$ Hui-Ching Lu, T. 2008, PRL, 101, 011301

\bibitem[Collett \& Auger (2014)]{Collett2014}Collett, T. E., $\&$ Auger, M. W. 2014, MNRAS, 443, 969

\bibitem[Collett (2015)]{Collett2015}Collett, T. E. 2015, ApJ, 811, 20

\bibitem[Conley et al. (2011)]{Conley2011}Conley, A., Guy, J., Sullivan, M., et al. 2011, ApJS, 192, 1

\bibitem[Eisenstein et al. (2005)]{Eisenstein2005}Eisenstein, D. J. et al. (SDSS Collaboration) 2005, ApJ, 633, 560

\bibitem[Eisenstein et al. (2011)]{Eisenstein2011}Eisenstein, D. J., Weinberg, D. H., Agol, E., et al. 2011, AJ, 142, 72

\bibitem[Ellis (2009)]{Ellis2009}Ellis, G. F. R., Gen. Relativ. Gravit. 2009, 41, 581

\bibitem[Etherington (1933)]{Etherington1933}Etherington, I. M. H., Philos. Mag. 1933, 15, 761; reprinted in Gen.Relativ. Gravit. 2007, 39, 1055

\bibitem[Fan et al. (2017)]{Fan2017}Fan, X.-L., Liao, K., Biesiada, M., Kurpas, A. P., $\&$ Zhu, Z.-H. 2017, PRL, 118, 091102

\bibitem[Faure et al. (2011)]{Faure2011}Faure, C., et al. 2011, A\&A, 529, A72

\bibitem[Futamase \& Yoshida (2001)]{Futamase2001}Futamase, T. \& Yoshida, S. 2001, Prog. Theor. Phys. 105, 5

\bibitem[Gavazzi et al. (2007)]{Gavazzi2007}Gavazzi, R., Treu, T., Rhodes, J. D., et al. 2007, ApJ, 667, 176

\bibitem[Gavazzi et al. (2008)]{Gavazzi2008}Gavazzi, R., Treu, T., Koopmans, L. V. E., et al. 2008, ApJ, 677, 1046

\bibitem[Gong \& Wang (2007)]{Gong2007}Gong, Y.-G., $\&$ Wang, A. 2007, PRD, 75, 043520

\bibitem[Grillo, Lombardi, \& Bertin (2008)]{Grillo2008}Grillo, C., Lombardi, M., \& Bertin, G. 2008, A\&A, 477, 397

\bibitem[Gurvtis. (1994)]{Gurvtis1994}Gurvits, L. I. 1994, ApJ, 425, 442

\bibitem[Gurvtis et al. (1999)]{Gurvtis1999}Gurvtis, L. I., Kellermann, K. I., $\&$ Frey, S. 1999, A$\&$A,  342, 378

\bibitem[Jiang \& Kochanek (2007)]{Jiang2007}Jiang, G. $\&$ Kochanek, C. S. 2007, ApJ, 671, 1568

\bibitem[Kessler \& Scolnic (2017)]{Kessler2017}Kessler, R., $\&$ Scolnic, D. 2017, ApJ, 836, 56

\bibitem[Koopmans \& Treu (2002)]{Koopmans2002}Koopmans, L. V. E. $\&$ Treu, T. 2002, ApJ, 568, L5

\bibitem[Koopmans \& Treu (2003)]{Koopmans2003}Koopmans, L. V. E., $\&$ Treu, T. 2003, ApJ, 583, 606

\bibitem[Koopmans et al. (2006)]{Koopmans2006}Koopmans, L. V. E., Treu, T., Bolton, A. S., Burles, S., $\&$ Moustakas, L. A. 2006, ApJ, 649, 599

\bibitem[Koopmans et al. (2009)]{Koopmans2009}Koopmans, L. V. E., et al. 2009, ApJL, 703, L51

\bibitem[Li, Fan \& Gou (2019)]{Li2019}Li, Y., Fan, X.-L., $\&$ Gou, L. 2019, ApJ in press, arXiv: 1901.10638

\bibitem[Li et al. (2014)]{Li2014}Li, Y.-L., Li, S.-Y., Zhang, T.-J., $\&$ Li, T.-P. 2014, ApJL, 789, L15

\bibitem[Li et al. (2011)]{Li2011}Li, Z.-X., et al. 2011, PLB, 695, 1

\bibitem[Li et al. (2012)]{Li2012}Li, Z.-X., et al. 2012, ApJ, 744, 176

\bibitem[Li et al. (2016)]{Li2016}Li, Z.-X., Wang, G.-J., Liao, K., $\&$ Zhu, Z.-H. 2016, ApJ, 833, 240

\bibitem[Li et al. (2018a)]{Li2018a}Li, Z.-X., Ding, X.-H., Wang, G.-J., Liao, K., $\&$ Zhu, Z.-H. 2018, ApJ, 854, 146

\bibitem[Li et al. (2018b)]{Li2018b}Li, Z.-X., Gao, H., Ding, X.-H., Wang, G.-J., $\&$ Zhang, B. 2018, Nat. Commun, 9, 3833

\bibitem[Liao et al. (2017a)]{Liao2017a}Liao, K., Fan, X.-L., Ding, X.-H., Biesiada, M., $\&$ Zhu, Z.-H. 2017, Nat. Commun, 8, 1148

\bibitem[Liao et al. (2017b)]{Liao2017b}Liao, K., Li, Z.-X., Wang, G.-J., Fan, X.-L. 2017, ApJ, 839, 70

\bibitem[Ma et al. (2019)]{Ma2019}Ma, Y., Cao, S., Zhang, J. et al. 2019, EPJC, 79, 121

\bibitem[Mackey et al. (2012)]{Mackey2012}Mackey, D. F., Hogg, D. W., Lang, D., $\&$ Goodman, J. 2012, PASP, 125, 306

\bibitem[Mario et al. (2019)]{Mario2019}Mario, H., et al. 2019, arXiv: 1906.04107

\bibitem[Marriner et al. (2011)]{Marriner2011}Marriner, J., Bernstein J. P., $\&$ Kessler, R., et al. 2011, ApJ, 740, 72

\bibitem[M\"ortsell \& J\"onsson (2011)]{Mortsell2011}M{\"o}rtsell, E., $\&$ J{\"o}nsson, J. 2011, arXiv:1102.4485

\bibitem[Newton et al. (2011)]{Newton2011}Newton, E. R., Marshall, P. J., Treu, T., et al. 2011, ApJ, 734, 104

\bibitem[Ofek et al. (2003)]{Ofek2003}Ofek, E. O., Rix H.-W., $\&$ Maoz, D. 2003, MNRAS, 343, 639

\bibitem[Oguri \& Marshall (2010)]{Oguri2010}Oguri, M., $\&$ Marshall, P. J. 2010, MNRAS, 405, 2579

\bibitem[Oguri et al. (2014)]{Oguri2014}Oguri, M., Rusu, C. E., $\&$ Falco, E. E. 2014, MNRAS, 439, 2494

\bibitem[Qi et al. (2018)]{Qi2018}Qi, J.-Z., et al. 2018, MNRAS, 483, 1104

\bibitem[R\"as\"anen et al. (2014)]{Rasanen2014}R{\"a}s{\"a}nen, S. 2014, JCAP, 03, 035

\bibitem[R\"as\"anen et al. (2015)]{Rasanen2015}R{\"a}s{\"a}nen, S., Bolejko, K., $\&$ Finoguenov, A. 2015, PRL, 115, 101301

\bibitem[Riess et al. (2019)]{Riess2019}Riess, A. G., et al. 2019, ApJ, 876, 85

\bibitem[Risaliti \& Lusso (2018)]{Risaliti2018}Risaliti, G., $\&$ Lusso, E. 2018, arXiv: 1811.02590

\bibitem[Ruff et al. (2011)]{Ruff2011}Ruff, A. J., et al. 2011, ApJ, 727, 96

\bibitem[Sapone et al. (2014)]{Sapone2014}Sapone, D., Majerotto, E., $\&$ Nesseris, S. 2014, PRD, 90, 023012

\bibitem[Schwab et al. (2010)]{Schwab2010}Schwab, J., Bolton, A. S., $\&$ Rappaport, S. A. 2010, ApJ, 708, 750

\bibitem[Scolnic et al. (2018)]{Scolnic2018}Scolnic, D. M., Jones, D. O., Rest, A., et al. 2018, ApJ, 859, 101

\bibitem[Seikel et al. (2012)]{Seikel2012}Seikel, M., Clarkson C., $\&$ Smith M. 2012, JCAP, 06, 036

\bibitem[Shafieloo \& Clarkson (2010)]{Shafieloo2010}Shafieloo, A. $\&$ Clarkson, C. 2010, PRD, 81, 083537

\bibitem[Shu et al. (2015)]{Shu2015}Shu, Y.-P., Bolton, A. S., Brownstein, J. R., et al. 2015, ApJ, 803, 71

\bibitem[Shu et al. (2016a)]{Shu2016a}Shu, Y., et al. 2016, ApJ, 824, 86

\bibitem[Shu et al. (2016b)]{Shu2016b}Shu, Y., et al. 2016, ApJ, 833, 264

\bibitem[Shu et al. (2017)]{Shu2017}Shu, Y.-P., Brownstein, J. R., Bolton, A. S., Burles, S., Koopmans, L. V. E., et al. 2017, ApJ, 851, 48

\bibitem[Sonnenfeld et al. (2013a)]{Sonnenfeld2013a}Sonnenfeld, A., Gavazzi, R., Suyu, S. H., Treu, T., $\&$ Marshall, P. J. 2013a, ApJ, 777, 97

\bibitem[Sonnenfeld et al. (2013b)]{Sonnenfeld2013b}Sonnenfeld, A., Treu, T., Gavazzi, R., et al. 2013b, ApJ, 777, 98

\bibitem[Sonnenfeld et al. (2015)]{Sonnenfeld2015}Sonnenfeld, A., et al. 2015, ApJ, 800, 94

\bibitem[Sullivan et al. (2011)]{Sullivan2011}Sullivan, M. et al. 2011, ApJ, 737, 102

\bibitem[Suyu et al. (2007)]{Suyu2007}Suyu, S. H., et al. 2007 AAS/AAPT Joint Meeting, American Astronomical Society Meeting 209, id.21.02; Bulletin of the American Astronomical Society, Vol. 38, p.927

\bibitem[Suzuki et al. (2012)]{Suzuki2012}Suzuki, N. et al. (Supernova Cosmology Project) 2012, ApJ, 746, 85

\bibitem[Tegmark et al. (2006)]{Tegmark2006}Tegmark, M. et al. (SDSS Collaboration) 2006, PRD, 74, 123507

\bibitem[Treu \& Koopmans (2002)]{Treu2002}Treu, T., $\&$ Koopmans, L. V. E. 2002, ApJ, 575, 87

\bibitem[Treu \& Koopmans (2004)]{Treu2004}Treu, T., $\&$ Koopmans, L. V. E. 2004, ApJ, 611, 739

\bibitem[Treu et al. (2006)]{Treu2006}Treu, T., Koopmans, L. V. E., Bolton, A. S., Burles, S., $\&$ Moustakas, L. A. 2006, ApJ, 640, 662

\bibitem[Treu et al. (2009)]{Treu2009}Treu, T., Gavazzi, R., Gorecki, A., et al. 2009, ApJ, 690, 670

\bibitem[Treu et al. (2010)]{Treu2010}Treu, T. 2010, Annu. Rev. Astro. Astrophys. 48, 87

\bibitem[Vegetti et al. (2010)]{Vegetti2010}Vegetti, S., Koopmans, L. V. E., Bolton, A., Treu, T., $\&$ Gavazzi, R. 2010, MNRAS, 408, 1969

\bibitem[Walsh et al. (1979)]{Walsh1979}Walsh, D., Carswell, R. F., $\&$ Weymann, R. J. 1979, Nature, 279, 381

\bibitem[Wong et al. (2015)]{Wong2015}Wong, K. C., Suyu, S. H., $\&$ Matsushita, S. 2015, ApJ, 811, 115

\bibitem[Wright et al. (2007)]{Wright2007}Wright, E. L. 2007, ApJ, 664, 633

\bibitem[Wei \& Wu (2017)]{Wei2017}Wei, J.-J., $\&$ Wu, X.-F. 2017, ApJ, 838, 2

\bibitem[Xia et al. (2017)]{Xia2017}Xia, J.-Q., Yu, H., Wang, G.-J., Tian, S.-X., Li, Z.-X., Cao, S., $\&$ Zhu, Z.-H. 2017, ApJ, 834, 75

\bibitem[Yu \& Wang (2016)]{Yu2016}Yu, H. $\&$ Wang, F.-Y. 2016, ApJ, 828, 85

\bibitem[Zhao et al. (2007)]{Zhao2007}Zhao, G.-B., Xia, J.-Q., Li, H., Tao, C., Virey, J. M., Zhu, Z.-H. $\&$ Zhang, X. 2007, PLB, 648, 8

\bibitem[Zhou \& Li (2019)]{Zhou2019}Zhou, H., $\&$ Li, Z.-X. 2019, CPC, 43, 3

\bibitem[Zhu \& Fujimoto (2002)]{Zhu2002}Zhu Z.-H., $\&$ Fujimoto M.-K. 2002, ApJ, 581, 1
\end{thebibliography}
\end{document}